\documentclass[aps,prb,twocolumn,groupedaddress,floatfix,letterpaper]{revtex4}

\usepackage{epsfig}
\usepackage{graphicx}
\begin{document}

\title{Optical characteristics of single wavelength-tunable InAs/InGaAsP/InP(100) quantum dots emitting at 1.55 $\mu$m}

\author{N.\ I.\ Cade}
\email{ncade@will.brl.ntt.co.jp}
\author{H.\ Gotoh}
\author{H.\ Kamada}
\author{H.\ Nakano}
\affiliation{NTT Basic Research Laboratories, NTT Corporation, Atsugi, 243-0198 Japan}

\author{S.\ Anantathanasarn}
\author{R.\ N\"{o}tzel}
\affiliation{eiTT/COBRA Inter-University Research Institute, Eindhoven University of Technology, 5600 MB Eindhoven, The Netherlands}

\date{\today}

\begin{abstract}
We have studied the emission properties of individual InAs quantum dots (QDs) grown in an InGaAsP matrix on InP(100) by metal-organic
vapor-phase epitaxy. Low-temperature microphotoluminescence spectroscopy shows emission from single QDs around 1550 nm with characteristic
exciton-biexciton behavior, and a biexciton antibinding energy of more than 2 meV. Temperature-dependent measurements reveal negligible
optical-phonon induced broadening of the exciton line up to 50 K, and emission from the exciton state clearly persists above 70 K.
Furthermore, we find no measurable polarized fine structure splitting of the exciton state within the experimental precision. These results
are encouraging for the development of a controllable photon source for fiber-based quantum information and cryptography systems.
\end{abstract}


\maketitle

There is currently considerable interest in the development of self-assembled quantum dot (QD) structures for novel telecommunication
applications, such as low-threshold lasers\cite{sugawara05} and non-classical light sources for quantum cryptography.\cite{michler00} In
the latter case, an optical fiber-based system requires the development of an efficient single-photon source operating in the fiber
transmission bands above 1260 nm. QD structures grown by metalorganic vapor-phase epitaxy (MOVPE) are very attractive commercially due to
the high growth rates achievable and the potential for monolithic integration into existing devices. Recently we have reported on the
photoluminescence (PL) characteristics of MOVPE grown InAs/InGaAs single QDs emitting at 1.3 $\mu$m.\cite{cade05} For longer wavelength
applications InAs/InP QDs are normally used; however, due to the small lattice mismatch careful control of the growth conditions is
required to realize emission around 1.55 $\mu$m.\cite{paranthoen01,kawaguchi04,pyun04} To date, there have been only a few investigations
into single QDs emitting in the important C-band region between 1.53--1.57 $\mu$m; these have used selective area chemical-beam
epitaxy,\cite{chithrani04} and very recently MOVPE techniques.\cite{takemoto04,saint06} However, in the latter cases the QDs were not
optimized for low temperature device applications, and suffered from a broad luminescence spectrum and low emission intensity at 1.55
$\mu$m relative to that at shorter wavelengths.

Here, we report on the emission properties of InAs QDs embedded in an InGaAsP matrix by MOVPE. Wavelength optimization has been achieved
via the insertion of ultra-thin GaAs interlayers. We present low-temperature PL spectra from a single QD with an emission wavelength around
1550 nm. Power-dependent measurements clearly reveal the formation of an exciton-biexciton system; the biexciton is found to be antibinding
with an emission energy of more than 2 meV relative to the exciton. The exciton linewidth shows negligible optical phonon induced
broadening up to 50 K. In addition, emission from discrete electronic states is seen clearly above 70 K, which suggests that these QDs may
be used as a single photon source operating at liquid-nitrogen temperatures.

The QD sample was grown at 500 $^{\circ}$C by low-pressure MOVPE on an InP (100) substrate misoriented 2$^{\circ}$ toward (110). A 100 nm
InP buffer layer and 100 nm lattice-matched InGaAsP layer ($\lambda_{Q} = 1.25$ $\mu$m) were deposited, followed by 2 monolayers (MLs) of
GaAs (growth rate 0.16 ML/s). The QDs were formed from 3 MLs of InAs, with a 5 second growth interruption and an upper 100 nm InGaAsP
layer. On top of this second InGaAsP layer, growth of the GaAs interlayer and InAs QDs was repeated at the same conditions for atomic force
microscopy (AFM) measurements; from these we obtain a QD sheet density of $\sim$$10^{10}$ cm$^{-2}$. The GaAs interlayer suppresses As/P
exchange during QD growth, thus reducing the QD height and blue-shifting the emission wavelength by a controllable amount. A detailed
description of the sample growth procedure and the macroscopic optical characteristics are published elsewhere.\cite{anant05}

To obtain single dot spectroscopy, mesa structures were fabricated by electron-beam lithography and dry etching with lateral sizes between
200 nm and 2 $\mu$m. Micro-PL measurements were taken using a continuous-wave (CW) Ar$^{+}$ laser (488 nm) focused to a $\sim$3 $\mu$m
spot; the luminescence was dispersed in a 0.5 m spectrometer and detected with a nitrogen cooled InGaAs photodiode array (instrument
resolution $\Gamma_{\textrm{\scriptsize{res}}} \simeq65$ $\mu$eV). The sample temperature was controlled using a continuous-flow He
cryostat.

\begin{figure}[tb!]
\epsfig{file=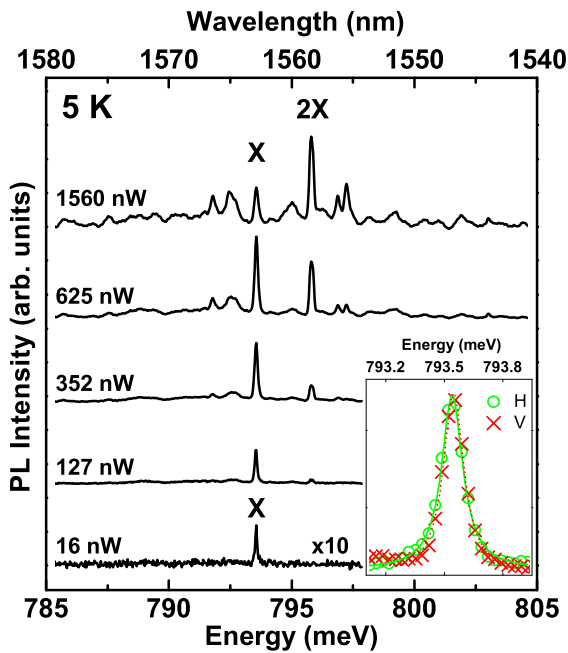,width=7.5cm} \caption{(Color online) PL spectra from a single QD in a 500 nm mesa, at different excitation powers.
Peaks X and 2X are attributed to neutral exciton and biexciton emission, respectively. (Inset) X emission resolved into horizontally (H)
and vertically (V) polarized components. The solid and dashed lines are Lorentzian fits.}\label{power}
\end{figure}

\begin{figure}[tb!]
\epsfig{file=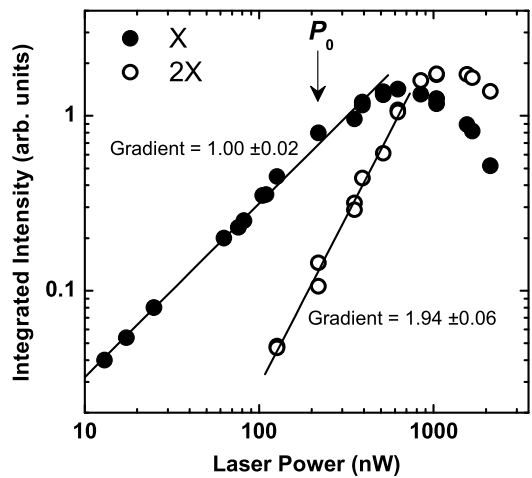,width=7cm} \caption{Integrated intensities of the X and 2X peaks in Fig.\ \ref{power}, as a function of CW laser
power. Solid lines are linear fits to the data.}\label{power2}
\end{figure}

PL spectra from a 500 nm mesa at 5 K are shown in Fig.\ \ref{power} for various excitation powers. At the lowest power there is a single
sharp emission line X in the spectral window between 1525--1580 nm. With increasing power additional lines appear in the spectrum; in
particular the line 2X develops superlinearly 2.3 meV above the emission energy of X. The lines X and 2X are attributed to recombination
from the neutral exciton and biexciton states, respectively, of a single QD. This assignment has been confirmed by plotting the integrated
intensities of these lines as a function of laser power, as shown in Fig.\ \ref{power2}: fits to the data give almost ideal linear and
quadratic behavior for the X and 2X lines respectively, which suggests a low scattering rate by impurities and defects for this particular
dot.\cite{nakayama95} The other spectral lines observed at higher powers most likely originate from charged- and multi-exciton
states.\cite{cade06} From a study of other QDs on the sample, we find similar exciton-biexciton behavior with 2X recombination energies in
the range 2-5 meV above the X line. This `antibinding' of the biexciton state has been observed previously in InAs/GaAs dots and results
from a reduction in exchange and correlation effects between the two localized excitons relative to the repulsive direct Coulomb
interaction.\cite{rodt03} This effect is consistent with the small dot aspect ratio (height/base diameter) of 0.09 expected from the growth
conditions.\cite{anant05}

The inset in Fig.\ \ref{power} shows the X line resolved into horizontally and vertically polarized components. Lorentzian fits to the data
suggest a fine-structure splitting of $<$10 $\mu$eV, which is smaller than the instrument precision and fitting error. This
indistinguishability is an important issue in the production of polarization entangled photon pairs for quantum information
applications.\cite{benson00} Furthermore, these dots show a very similar emission intensity under pulsed (80 MHz) excitation from a
Ti:sapphire laser which is necessary for controlled generation of single photons.

\begin{figure}[tb!]
\epsfig{file=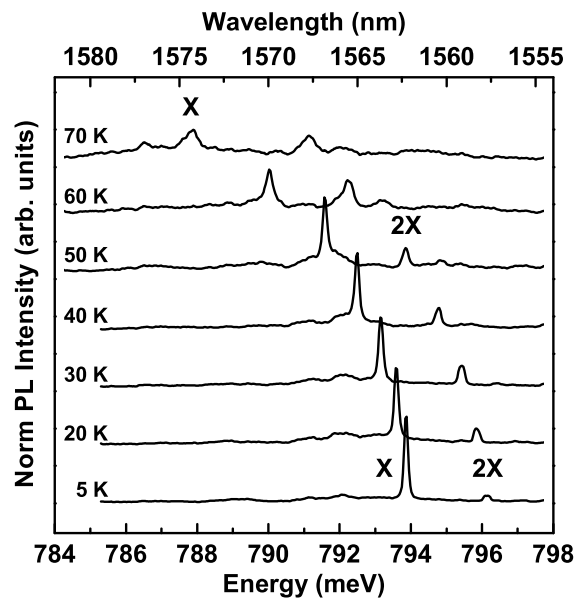,width=7.5cm} \caption{PL spectra from the same QD as in Fig.\ \ref{power}, normalized to the X integrated intensity,
as a function of temperature. The laser power was $P_{0}$, indicated in Fig.\ \ref{power2}.}\label{temp}
\end{figure}

\begin{figure}[tb!]
\epsfig{file=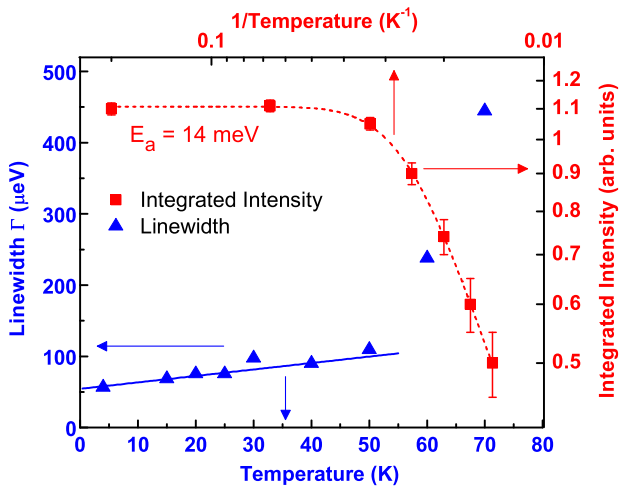,width=8.2cm} \caption{(Color online)(Bottom-left axes) Temperature dependence of the corrected exciton PL linewidth
$\Gamma$ (triangles). The solid line is a linear fit over low temperatures. (Top-right axes) Temperature dependence of the exciton PL
integrated intensity. A data fit (dashed line) gives an activation energy of 14 meV. Note, the two horizontal axes do not correspond
exactly.}\label{temp2}
\end{figure}

Figure \ref{temp} shows temperature-dependent PL spectra from the same QD studied in Fig.\ \ref{power}, normalized to the integrated
intensity of the X line. The exciton emission intensity and linewidth\cite{note1} determined from these spectra are plotted in Fig.\
\ref{temp2}; the exciton line appears thermally stable over the measured temperature range, with the intensity at 70 K only dropping to
approximately half of the maximum value. A fit to the data gives a thermal activation energy of 14 meV. We have observed similar behavior
in other QDs on the sample, with well resolved emission from the exciton state at 77 K.

At low temperatures the exciton-optical phonon interaction is negligible and the X linewidth $\Gamma$ has a linear temperature dependence
due to acoustic phonon scattering: $\Gamma(T)=\Gamma_{0}+\alpha T$, where $\Gamma_{0}$ is the linewidth at 0 K. A linear fit of the data in
Fig.\ \ref{temp2} gives $\alpha=0.9$ $\pm$0.2 $\mu$eV/K, and $\Gamma_{0}\simeq 50$ $\mu$eV. The value of $\alpha$ is similar to those
previously reported for other QD systems,\cite{besombes01,kammerer02} and significantly smaller than that of a quantum well system due to
the absence of final states for scattering. Above 50 K there is a sharp increase in linewidth due to optical phonon scattering, and the
line shape strongly deviates from a Lorentzian profile.\cite{besombes01} From different QDs we find similar values for $\alpha$, but large
variations in $\Gamma_{0}$; this latter effect is most likely due to the influence of charge fluctuations on the mesa sidewalls when using
nonresonant laser excitation.\cite{bayer02a}

In conclusion, we have studied the emission properties of individual InAs/InGaAsP QDs grown on InP(100) by MOVPE; the insertion of a thin
GaAs interlayer has enabled tuning of the QD emission wavelength to 1.55 $\mu$m for telecom applications. We observe almost ideal
exciton-biexciton behavior at low temperatures, with a biexciton antibinding energy of more than 2 meV. The exciton line shows negligible
broadening from optical phonon scattering up to 50 K and appears thermally stable at higher temperatures, with clearly resolvable emission
above 70 K. Furthermore, there is no measurable fine structure splitting within the experimental precision. These results suggest that QDs
fabricated with this growth technique may be suitable as an on-demand single photon source at liquid nitrogen temperatures, for fiber-based
quantum information and cryptography systems.

The authors are grateful to T.\ Segawa at NTT Photonics Laboratories for etching the mesa structures. This work was partly supported by the
National Institute of Information and Communications Technology (NICT).


\begin{thebibliography}{}
\bibitem{note1} We plot the corrected linewidth $\Gamma=(\Gamma^{2}_{\textrm{\scriptsize{exp}}}-\Gamma^{2}_{\textrm{\scriptsize{res}}})^{1/2}$,
 where $\Gamma_{\textrm{\scriptsize{exp}}}$ is the measured linewidth and $\Gamma_{\textrm{\scriptsize{res}}}$ the instrument resolution.
\end{thebibliography}
\end{document}